\documentstyle[twoside,fleqn,psfig,epsf,espcrc2]{article}%
\begin{document}                                                        
\renewcommand{\refname}{\normalsize\bf References}
\title{%
One-parameter superscaling in three dimensions
}

\author{%
        J\'anos Pipek%
        \address{Elm\'eleti Fizika Tansz\'ek, 
                 Budapesti M\H uszaki \'es Gazdasagtudom\'anyi Egyetem,\\
                 H-1521 Budapest, Hungary}%
        \, Imre Varga$^{\rm a,}$%
        \address{Fachbereich Physik, Philipps--Universit\"at Marburg,\\
                 Renthof 6, D-35032 Marburg an der Lahn, Germany}%
\thanks{Supported by OTKA T029813, T032116 and F024135 and the Alexander
        von Humboldt foundation}
        \, and Etienne Hofstetter
        \address{Management School, Imperial College, London SW7 2BX,
                 United Kingdom}
}
%
%
\begin{abstract}
\hrule
\mbox{}\\[-0.2cm]

\noindent{\bf Abstract}\\

Numerical and analytical details are presented on the
newly discovered superscaling property of the energy spacing 
distribution in the three dimensional Anderson model. 
\\[0.2cm]
\hrule
\end{abstract}

\maketitle

\section{Introduction}

The statistical properties of the spectra of disordered systems
continues to be a vivid field of research both analytically and
numerically. It is by now well established that the spectral
fluctuations in the limit of vanishing disorder are described by the
random matrix theory (RMT) \cite{Mehta,PhRep} with corrections of 
the order of $1/g$, where
$g$ is the dimensionless classical conductance of the system. This
is because in this regime the states are extended (ergodic). In
the limit of infinite disorder, on the other hand, due to the extremely
strong localization of the eigenstates the levels are practically
uncorrelated at least in the thermodynamic limit. 

The first attempts \cite{Shkl} to investigate the statistics at the 
localization--delocalization transition (LDT) (also known as the
metal--insulator transition), showed that indeed the statistics at the LDT 
is different from the two other extremes. Eventually it shares some
of the features of both of them. It has been shown numerically in
many simulations \cite{Shkl,GOE,GUE,GSE,Daniel,IV} 
that e.g. the nearest neighbor spacing distribution,
$P(s)$, the probability of finding no eigenlevels within an energy window 
of length $s$, behaves for low--$s$ as $P(s)\sim s^{\beta}$, and
for large--$s$ as $\ln P(s)\sim -as$, where $\beta=1,2$ and 4 
depending on the global orthogonal, unitary and symplectic symmetry of
the system, and $a$ is a constant that depends weakly on $\beta$
but strongly on dimensionality $d$ \cite{GOE,GUE,GSE}. The presence of 
the level repulsion at low--$s$ is a direct consequence that the 
eigenstates at the LDT are heavily fluctuating but still strongly 
overlapping \cite{FM}. The heavy fluctuations expressed as their 
multifractal character \cite{Martin} enters in the parameter $a$ that 
is connected both to the correlation dimension of the states and the 
level compressibility \cite{Chalker}. Such a relation shows how the 
statistical properties of the levels and the states are coupled at the LDT.

In a previous publication \cite{IV} we have presented numerical 
evidence that using appropriate indices to describe the overall 
shape of $P(s)$, a superuniversal (i.e. $\beta$ independent) scaling 
relation is revealed as the system undergoes the transition from the 
metal to an insulator. Here we give further evidence that with the change of
system size $L$ and disorder $W$ the nearest neighbor
spacing distribution, $P(s)$, evolves as a one--parameter family of
distributions. A similar evolution has been shown analytically to exist by 
Shapiro \cite{Boris} for the conductance distribution in $d=2+\varepsilon$. 
The parameter in our case is also the scaling variable that can be chosen
as $g$, i.e. the classical, dimensionless conductance of the system.

\section{The model and the analysis}

In order to study the statistical properties of disordered systems with
and without time reversal symmetry and with broken rotational symmetry
we have used the Anderson model which is a tight--binding model defined 
on a $L^3$ lattice. For the details we refer to the review of Kramer and
MacKinnon \cite{Ang} and also our previous work \cite{IV}. In Ref. \cite{IV}
the numerical results have been illustrated using the symplectic case
where the rotational symmetry has been broken with the introduction of 
spin--orbit scattering. Here we present the results of the case with
(without) time reversal symmetry corresponding to the orthogonal (unitary)
symmetry. Time reversal symmetry has been broken by the introduction
of a magnetic field. 
\begin{figure}[ht]
\vspace{-0.3in}
\leavevmode
\epsfxsize=7cm
\epsfbox{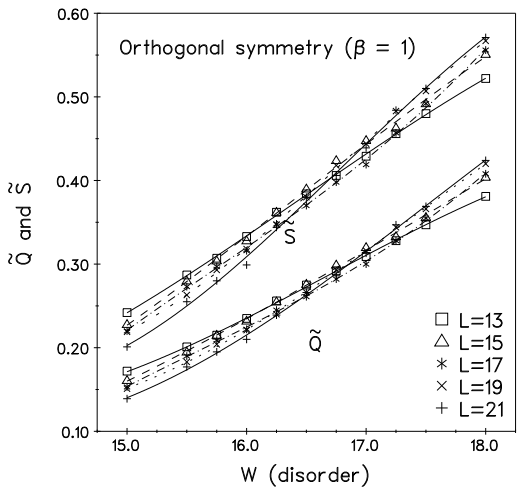}
\vspace{-0.5cm}
\caption{\label{b1} $\tilde Q(L,W)$ and $\tilde S(L,W)$ for
    the case of orthogonal symmetry. Continuous curves are polynomial
    fits.}
\vspace{-0.5cm}
\end{figure}
\begin{figure}[ht]
\vspace{-0.3in}
\leavevmode
\epsfxsize=7cm
\epsfbox{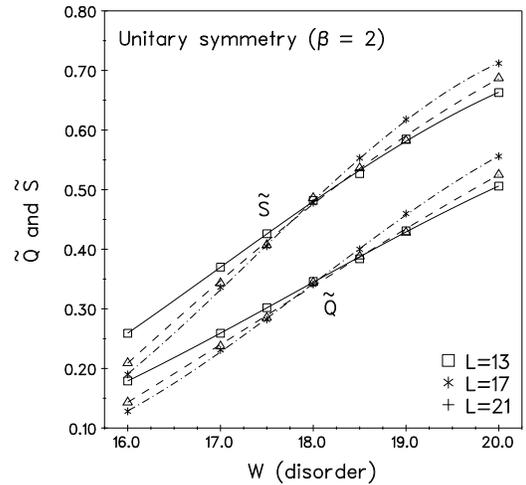}
\vspace{-0.5cm}
\caption{\label{b2} $\tilde Q(L,W)$ and $\tilde S(L,W)$ for
    the case of unitary symmetry. Continuous curves are polynomial
    fits.}
\vspace{-0.5cm}
\end{figure}

After appropriately unfolding the spectra in many samples of the system
we calculated the nearest neighbor spacing distribution, $P(s)$. For
each system size $L$ and disorder $W$ the number of spacings was of the order
of $10^5$. In order to characterize the shape of $P(s)$ we use generalized 
R\'enyi--entropies \cite{Imr}
\begin{equation}
\label{qsstr}
    q=\mu_2^{-1}\quad \hbox{and}\quad S_{str}=\mu_S +\ln \mu_2,
\end{equation}
where  $\mu_2=\langle s^{2}\rangle$ is the second moment of $P(s)$, 
while $\mu_S=-\langle s \ln s \rangle$. After a proper rescaling we
obtain
\begin{eqnarray}
    \label{qtilde}
    -\ln (q)&\to &
    \frac{-\ln (q)+\ln (q_W)}{-\ln (q_P)+\ln (q_W)}=\tilde Q \nonumber \\ 
    \label{stilde}
    S_{str}&\to &
    \frac{S_{str}-S_W}{S_P-S_W}=\tilde S
\end{eqnarray}
where index $_{P}$ refers to the Poissonian, uncorrelated case 
and $_{W}$ to the Wigner--surmise representing the RMT cases respectively.
Their values are listed in Table~\ref{t1}. Such a rescaling ensures 
that $\tilde S=\tilde Q=0(1)$ belonging to the RMT (Poisson) limit.
\begin{table}[ht]
\caption[params]{\label{t1} Shape descriptive parameters for the case
    of different $P(s)$ functions.}
\begin{center}
\begin{tabular}{crrrrr}
          &  Poisson &  $\beta =1$   & $\beta=2$  &  $\beta=4$  \\
\hline
$q$       & 0.5295   & 0.7854 & 0.8488 & 0.9054 \\
$-\ln(q)$ & 0.6358   & 0.2416 & 0.1639 & 0.0994 \\
$S_{str}$ & 0.2367   & 0.1025 & 0.0733 & 0.0464 \\
\end{tabular}
\end{center}\vspace{-0.3in}
\end{table}
In Figs.~\ref{b1} and \ref{b2} we present the way how $\tilde S$ and 
$\tilde Q$ vary as a function of disorder for various system sizes. 
The scale invariance at around $W=W_c$ shows the presence of the LDT.

In Table \ref{t2} we collected the values of the critical disorder and the 
critical exponent obtained from a linearization around this fixed point.
Even though the values presented in Table \ref{t2} agree well with the 
ones known from earlier calculations, the precision 
is somewhat lower. 

\begin{figure}[ht]
\vspace{-0.3in}
\leavevmode
\epsfxsize=7cm
\epsfbox{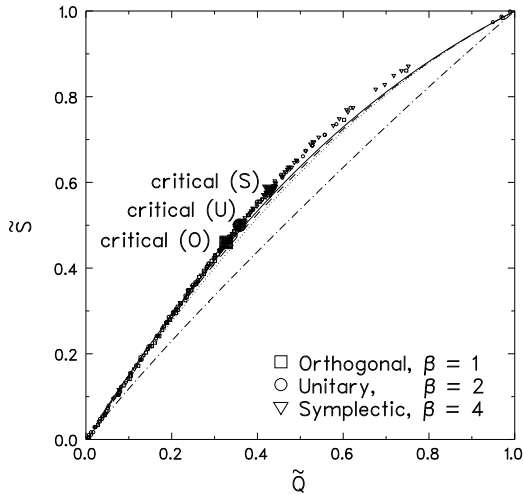}
\vspace{-0.5cm}
\caption{\label{all} 
$\tilde S(L,W)$ as a function of $\tilde Q(L,W)$ for all the symmetry 
classes. The solid symbols represent the positions of the critical points.
The continuous (solid, dashed, dotted) curves are our analytical estimates, 
see text for details. The dashed dotted line shows the relation of the 
$P(s)$ due to Izrailev~\protect\cite{izr} representing the pure effect 
of level repulsion.}
\vspace{-0.5cm}
\end{figure}

In Figure~\ref{all} we plot the $\tilde S$ as a function of $\tilde Q$
for all the symmetries ($\beta=1,2$ and 4). As a comparison we also
plotted our analytical estimates based on a phenomenological assumption
and the relation obtained for the interpolating formula of Israilev 
\cite{izr}. 
\begin{table}[ht]
\caption[params]{\label{t2} Position of the critical point and
    critical exponents obtained using the shape descriptive parameters
    $\tilde Q$ and $\tilde S$.}
\begin{center}
\begin{tabular}{crr}
               & $W_c$\quad\     & $\nu$\qquad\   \\
\hline
$\beta =1$ (O) & 16.77$\pm $0.63 & 1.30$\pm $0.38 \\
$\beta =2$ (U) & 18.13$\pm $0.38 & 1.63$\pm $0.57 \\
$\beta =4$ (S) & 21.97$\pm $0.17 & 1.41$\pm $0.32 \\
\end{tabular}
\end{center}\vspace{-0.3in}
\end{table}
We will show below evidence that the overall scaling behavior in
Figure~\ref{all} can be understood if the spacing distribution for
any finite value of $W$ and $L$, i.e. for $0<g<\infty$ is given as
\begin{equation}
    {\cal P}_{g,\beta}(e^x)=
    \int_{-\infty}^{\infty }{\cal Q}_{g,\beta}(e^{x-y})
    {\cal W}_{\beta}(e^y)dy
\label{conv}
\end{equation}
with $e^x\equiv s$. Here the function ${\cal W}_{\beta}(t)$ is simply
the Wigner--surmise and ${\cal P}_{g,\beta}(s)$ is the histogram
measured in the numerical simulation. All three functions, $\cal P,Q$ 
and $\cal W$ are spacing distribution functions. 

In the next section we will prove that for the convolution of the type
(\ref{conv}) the R\'enyi--entropies introduced in \cite{Imr} are additive.
Therefore, apart from the denominator, the constant shift in 
(\ref{stilde}) corresponds to ${\cal W}_{\beta}$ in (\ref{conv}),
hence we may conclude that the relation
in Figure~\ref{all} is described by the properties of function 
$\cal Q$.

We will also
show that the solution of (\ref{conv}) for ${\cal Q}_{g,\beta}(u)$ in the
extreme cases ($g=0$ and $g\to\infty$) can be done analytically and
that a phenomenological interpolation can describe the $\tilde S$ vs.
$\tilde Q$ relation in Figure~\ref{all}.

\section{Additivity of R\'enyi entropies}

The expression (\ref{conv}) is in fact a convolution and after a variable
transformation can be written as
\begin{equation}
    {\cal P}_{g,\beta}(s)=
    \int_{0}^{\infty}{\cal Q}_{g,\beta}\left (\frac{s}{t}\right )
    {\cal W}_{\beta}(t)\frac{dt}{t}.
\label{conv2}
\end{equation}
Let us denote the $k$-th moment of a distribution $\cal F$ by 
$\mu_k[{\cal F}]$. Then it is easy to see from (\ref{conv2}) (omitting
indices $g$ and $\beta$) that
\begin{eqnarray}
    \mu_k[{\cal P}]&=&\int_0^{\infty}ds\,s^k{\cal P}(s) \nonumber \\
                   &=&\int_0^{\infty}\frac{dt}{t}{\cal W}(t)
                      \int_0^{\infty}ds\,s^k{\cal Q}
                                     \left (\frac{s}{t}\right ) \nonumber \\
                   &=&\int_0^{\infty}dt\,t^k{\cal W}(t)
                      \int_0^{\infty}du\,u^k{\cal Q}(u) \nonumber \\
                   &=&\mu_k[{\cal W}]\mu_k[{\cal Q}],
\end{eqnarray}
which for $q=-\ln\mu_2$ yields the desired additivity. 
A similar procedure
yields the additivity of the quantity $S_{str}$. 
The key step is to
realize that the entropic moment of $\cal F$ is $\mu_S[{\cal F}]=
\langle s\ln s\rangle$ which can be expressed as
\begin{equation}
\mu_S[{\cal P}]=\mu_1[{\cal Q}]\mu_S[{\cal W}]+\mu_1[{\cal W}]\mu_S[{\cal Q}].
\end{equation}

\section{Phenomenological approximation}

Here we show that the solution of (\ref{conv}) for ${\cal Q}_{g,\beta}(u)$ in 
the extreme cases ($g=0$ and $g\to\infty$) can be obtained analytically and
that a phenomenological interpolation can describe the $\tilde S$ vs.
$\tilde Q$ relation in Figure~\ref{all}. First it is easy to see that 
for the case of vanishing disorder ($g\to\infty$) we expect that 
${\cal P}(s)\to{\cal W}(s)$. In this case ${\cal Q}(u)=\delta (u-1)$. 
On the other hand at the other extreme, when $g\to 0$, another universality
is expected: ${\cal P}(s)\to \exp(-s)$, the
Poissonian distribution. In this case after nontrivial calculations, 
Eq.~(\ref{conv2}) is solved
for ${\cal Q}_{0,\beta}(s)\equiv R(bs)$
\begin{equation}
    R(y)=a\left \{
    \begin{array}{ll}
        e^{-y^2}                 & {\rm (O)} \\
        {\rm erfc}(y)                 & {\rm (U)} \\
        (2y^2+1){\rm erfc}(y)-\frac{2y}{\sqrt{\pi}}e^{-y^2}
                                 & {\rm (S)}
    \end{array}\right.
\label{wp}
\end{equation}
\noindent
where $y=bs$, with $b=1/\sqrt{\pi}$, $\sqrt{\pi}/4$, and $3\sqrt{\pi}/16$
and $a=2/\pi$, $\pi/4$, and $9\pi/32 $ for $\beta=1$, $2$ and $4$, 
respectively. The complementary error function is denoted as erfc.
Now a phenomenological step is introduced in order to describe the
evolution of ${\cal P}$ for any finite value of $g$. We propose a simple
form
\begin{equation}
    {\cal Q}_{g,\beta}(s)= a_{g,\beta}s^gR(b_{g,\beta}s)
\label{interp}
\end{equation}
where the coefficients $a_{g,\beta}$ and $b_{g,\beta}$ are determined
from the normalization conditions $\mu_0[{\cal Q}]=\mu_1[{\cal Q}]=1$.
Obviously for $g\to 0$ they yield the values given above for (\ref{wp}).
For $g\to\infty$ ${\cal Q}(s)\to\delta (s-1)$, as well. The $\tilde S$ vs. 
$\tilde Q$ relation for this interpolating function is depicted as
continuous curves in Fig.~\ref{all}. They still have a small $\beta$ 
dependence but follow the data very closely. The $\beta$--dependence can 
be attributed to ($i$) the simple interpolation we used in (\ref{interp})
and to ($ii$) the fact that we focused only on the linear shift introduced
in $(\ref{stilde})$ and disregarded the denominator.

\section{Conclusion}

In this paper we have given further details, both numerical and analytical
of a recently discovered scaling relation existing in the evolution of
the nearest neighbor spacing distribution, $P(s)$, as the dimensionless
conductance $g$ changes from 0 (localized) to $\infty$ (metal).

\end{document}